\documentclass[prl,english,twocolumn,amsmath,showpacs]{revtex4}
\usepackage{graphicx}
\usepackage{color}
\usepackage{babel}

\begin{document}

\title{Surface Spin-valve Effect}

\author{I. K. Yanson$^{1}$, Yu. G. Naidyuk$^{1,2}$, V. V. Fisun$^{1,2}$,
A. Konovalenko$^{2}$,
O. P. Balkashin$^{1}$, L. Yu. Triputen$^{1}$, and V. Korenivski$^{2}$}

\affiliation{$^{1}$B. Verkin Institute for Low Temperature Physics and Engineering,
National Academy of Sciences of Ukraine, 47 Lenin Avenue, 61103 Kharkiv,
Ukraine, \\
$^{2}$Nanostructure Physics, Royal Institute of Technology,
SE-10691 Stockholm}

\date{\today{}}

\begin{abstract}
We report an observation of spin-valve like hysteresis within a few
atomic layers at a ferromagnetic interface. We use phonon spectroscopy
of nanometer sized point contacts as an \emph{in-situ} probe to study
the mechanism of the effect. Distinct in energy phonon peaks for
contacts with dissimilar nonmagnetic outer electrodes allow to localize
the observed spin switching to the top or bottom interfaces for nanometer
thin ferromagnetic layers.
The mechanism consistent with our data is energetically distinct atomically
thin surface spin layers that can form current or field driven \emph{surface
spin-valves} within a \emph{single} ferromagnetic film.

\pacs{72.10.Di, 72.25Mk, 73.40Jn}
\end{abstract}
\maketitle

Spin-valves in the current perpendicular to the plane geometry are
usually nanopillars having two ferromagnets (F1 and F2) of different
anisotropy, such that one is magnetically hard and the other is magnetically
soft separated by a nonmagnetic spacer layer (N). The conductivity
of such a spin-valve is governed by the giant magnetoresistance effect
\cite{Fert,Dieny} and depends on the mutual orientation of the magnetization
in F1 and F2 \cite{Katine}. Recent studies have shown \cite{Meyers,Ji,Ozy,Yanson}
that nonmagnetic metal contacts to single ferromagnetic films (N-F)
exhibit spin torque effects similar to those observed in F1/N/F2 spin-valves.
For single N-F interfaces, non-hysteretic singularities of magnetic
origin observed in the conductance are
explained as arising from spin wave excitations in the ferromagnetic
film \cite{Polianski,Stiles}. Similar to the nanopillar case \cite{Katine},
these peaks are observed only for one polarity of the bias current,
namely for the electron current flowing from N into F, and their position
on the current/voltage axis is proportional to the magnitude of the
external magnetic field \cite{Ji,Ozy,Yanson}. Another pronounced
and rather unexpected feature of N-F nano-contacts, being the trade
mark of the F1-N-F2 spin-valves, is hysteresis in resistance versus
voltage, resulting in a bistable resistance state near zero bias.
Its origin is under debate, with proposed interpretations ranging
from surface exchange anisotropy \cite{Chen} and magneto-elastic
anisotropy \cite{Kono} to spin vortex states \cite{ozy1}. In this
work we investigate the mechanism of the hysteretic conductance in
magnetic point contacts (PC's) by combining PC phonon spectroscopy
\cite{book} and magnetoconductance on the scale down to $\sim$1
nm in the contact radius, which is inaccessible by today's lithographic
means. We focus in particular on measuring thin films, where the magnetic
film thickness ($t$) can be smaller than the PC diameter ($d$) and
comparable or smaller than the ferromagnetic exchange length in the
material. Such experimental configuration prevents formation of volume
domains on the scale of the contact. Furthermore, we investigate the
regime of dimensional cross over in $t$ vs. $d$, the latter scale
defining the contact core where the current density is maximum. In
the limit $t<d$ both magnetic interfaces can be
inside the high current density region and therefore actively contribute
to spin transport, whereas for $t>d$ only one interface is expected
to contribute to magnetoconductance. Such nanometer scale probing
into the ferromagnetic surface, with an in-situ spectroscopic detection
of the location of the nano-contact core, leads us to conclude that
the observed magnetic effects are due to atomically thin surface spin
layers acting as current or field driven spin-valves with respect
to the magnetization in the interior of the ferromagnetic layer. Atomic
scale domain walls of this kind have recently been shown to explain
unusual surface magnetism in Co and Fe \cite{Gruyters,Jonker}. The
surface spin valve effects we report should be of quite general character,
present in all spintronic devices provided the effect is not averaged
out by interface imperfections on going from the nanometer scale to
nanopillar structures of tens and hundreds of nanometers in size.

Our samples are Co films deposited onto oxidized Si substrates buffered
with a non-magnetic bottom electrode (N2=Cu, Au) of 30 to 100 nm in
thickness. The thickness of the Co films (F) ranged between 2 and
100 nm. For several samples the Co layer was capped with a 3 nm Au
layer in order to prevent oxidation of the Co surface. The other non-magnetic
electrode (N1=Cu, Ag, Au, W, Mo) is prepared in the form of a sharp
tip, mechanically manipulated at low temperature to gently touch the
F surface. All measurements are done at 4.2 K, with the samples always
cooled from room temperature in zero field. The magnetic field is
applied parallel to the film plane. For each contact a set of complementing
transport characteristics were recorded: the differential resistance
$R(V)\equiv dV/dI(V)$, magnetoresistance $R(H,V\approx0)$, and the
so called PC spectrum, $d^{2}V/dI^{2}(V)$, provided
the contact was mechanically stable for a sufficient period of time.
Great many shorter lived contacts have been measured, resulting in
a large library of still very informative subsets of magnetotransport
data. The static resistance measured at low bias showed essentially
the same behavior as the differential resistance. In the discussion
to follow we therefore make no distinction between the two and restrict
the data shown to $dV/dI$.

\begin{figure}[t]
\begin{center}
\includegraphics[width=4cm,angle=0]{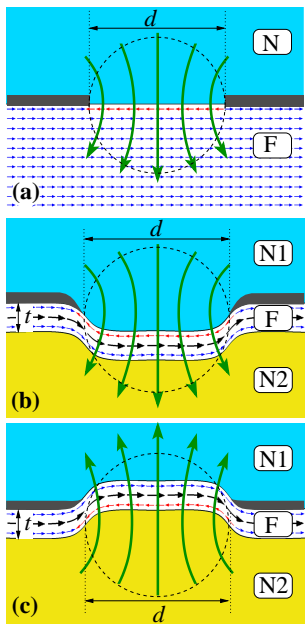}
\vspace{-0cm}
\end{center}
\caption{ Schematic of three generic point contacts: (a) a nonmagnetic
tip in contact with a bulk or thick film ferromagnet ($t\gg d$);
a thin F layer ($t<d$) between two nonmagnetic metals N1 and N2,
placed in the lower (b) or upper (c) part the contact core. The
electrodes are separated by a non-conductive surface layer, shown
in dark grey. Green arrows indicate the electron current flow through
the PC core, which is schematically shown as a thin dashed circle.
Small red arrows illustrate magnetic switching of the interface spins
with respect to the interior spins (black arrows). Notice that the
polarity of the electron current in (b) and (c) are opposite,
illustrating the cases of maximum current density and hence strongest
spin torques at the top and bottom F surface, respectively}
\label{f1}
\end{figure}

Three generic geometries of a point contact are illustrated in Fig.\,1. 
Fig.\,1a shows a schematic view of a nonmagnetic metal (N) in contact
with a bulk or a thick film ferromagnet. The two metals are separated
electrically except for a small circular orifice of diameter $d$,
the point contact size. This model corresponds to the experimental
configuration studied previously \cite{Ji,Chen,Yanson}. In this case,
only the upper interface between N and F is located in the region
of high current density, hence only this interface plays a role in
magneto-transport.

Fig.\,1b illustrates the case where a thin F film is located in the
lower half of the contact core (dashed circle). The upper F-N interface
is located in the region of maximal current density and therefore
contributes most to the spin torque effects. Notice that the contact
core in this case is filled predominantly with N1 material, which
therefore should dominate the PC spectrum. For
a Co layer thinner than $d$ its contribution to the PC spectrum is
expected to be small \cite{book}. The Co phonon peaks are therefore
not seen in the PC spectra discussed below.

A third characteristic, albeit less probable geometry is shown in
Fig.\,1c. Here, due to a supposed protrusion on the surface of N2,
the Co nanolayer is elevated to the upper half of the contact core
filled predominantly by N2, which therefore is expected to dominate
the phonon PC spectrum. Consequently, the maximal current density
is found at the bottom interface of the Co film. Thus, using PC phonon
spectroscopy we can determine \emph{in-situ} the relative weight of
the two N/F interfaces in the magneto-transport for a given contact.

Fig.\,2a illustrates the \emph{normal} hysteresis for a hetero-contact
Cu100/Co3-Au, a 100 nm Cu buffer electrode covered with a 3 nm thick
Co film in contact with an Au tip. The higher resistance state is
obtained for negative bias where the electrons flow from N1 to F.
Cycling through a large positive bias (red curve), in this case approximately
+40 mV, switches the contact into the low resistance state at about
+10 mV. This state is preserved down to about -40 mV, followed by
a smeared upward transition. Assuming the hysteresis is caused by
a domain wall of type F1-DW-F2 within the Co layer, the large negative
bias produces an anti-parallel (AP, or at least a large relative angle)
configuration of the F1 and F2 magnetic sublayers, while for the large
positive bias the spins in the two sublayers are parallel (P). The
subsequently recorded PC spectrum shown in Fig.\,2b, which is the measure
of the electron-phonon interaction in the contact \cite{book}, shows
that the contact core is occupied predominantly by Au (N1). This means
that the Co layer (whose phonon spectral lines are not detected owing
to its small thickness) is located in the lower half of the PC core
and, therefore, it is the upper N/F interface that plays the main
role in the spin torque driven hysteresis.

\begin{figure}[t]
\begin{center}
\includegraphics[width=9cm,angle=0]{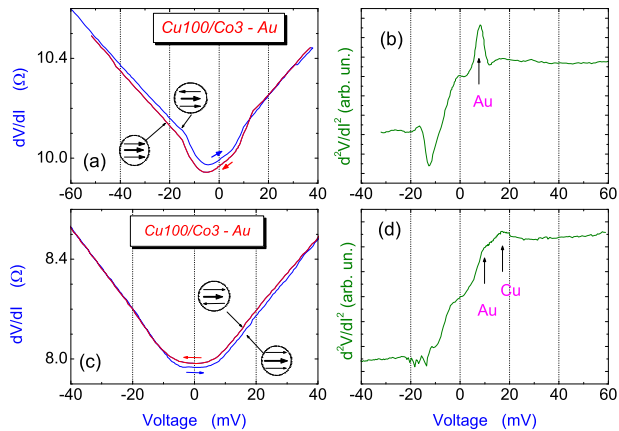}
\vspace{-1cm}
\end{center}
\caption{ (a) \emph{Normal} hysteresis in $dV/dI(V)$ for a Cu100/Co3-Au
contact; (b) PC spectrum for the same contact showing a pronounced
Au transverse phonon peak; (c) \emph{anomalous} hysteresis in $dV/dI(V)$
for a nominally similar contact (Cu100/Co3-Au), along with its PC
spectrum showing a dominant Cu phonon maximum - (d). The arrows in
dashed circles in (a,c) schematically indicate the orientation of
the surface and interior spins. The red and blue arrows in (a,c) as 
well as in Figures 3 and 4 indicate the bias sweep direction.}
\label{f2}
\end{figure}

The typical resistance of our contacts is $\sim10$ $\Omega$, which
correspond to the Sharvin diameter $d\simeq$10 nm. This is estimated
as follows (see Fig.\,3.9 and the accompanying text in \cite{book}
for additional deatails): \[
R_{\text{Sharvin}}=16\rho l/3\pi d^{2},\]
which using the free electron approximation and the Fermi momentum
$k_{\text{F}}(\textrm{Cu,Au)}\simeq1.35\cdot10^{8}\textrm{cm}^{-1}$
yields \[
d\ [\text{nm}]\simeq30/\sqrt{R\ [\Omega]}.\]
This value somewhat underestimates the true contact diameter since
the Sharvin formula applies in the ballistic current regime. For the
contacts reported herein the regime is closer to diffusive (the electron
mean free path $l<d$), hence the typical contact diameter should
be $\gtrsim10$ nm for $R\approx10\ \Omega$. This means that for
our nanometer thin ferromagnetic layers the condition $t<d$ is fulfilled
for typical contacts, hence both ferromagnetic interfaces can be expected
to contribute to magnetotransport, depending on the microscopic layout
of the contact (see Fig.\,1b,c).

Fig.\,2c illustrates the case where the bottom interface is dominant,
which leads to a reversal of the hysteresis in conductance. Such inverse
or anomalous hysteresis is observed much less frequently owing to
the fact that protrusive point contacts (see Fig.\,1c) are less probable
micromechanically. The outer electrodes were chosen with an aim to
separate their main phonon peaks, which in the case of Cu and Au are
found at approximately 17 and 10 mV, respectively \cite{book}. The
separation of 7 mV is large enough to reliably distinguish the peaks
even for non-ballistic contacts with smeared phonon maxima \cite{Yanson}.
The dominant Cu maximum in Fig.\,2d, in contrast to the spectrum of
Fig.\,2b, indicates that it is the bottom Co/Cu interface region that
plays the main role in the electron and spin transport for this nano-contact.
The corresponding hysteresis is \emph{anomalous}, i.e., the large
resistance state corresponds to the electrons flowing from the film
into the tip (Fig.\,2c). The spectroscopic data provides a natural
explanation, namely, the contact core of highest current density is
at the bottom F/N interface where the electrons flow from the Cu bottom
electrode into Co. The transition to the high resistance state occurs
above +35 mV and the reverse transition at -8 mV.

\begin{figure}[t]
\begin{center}
\includegraphics[width=9cm,angle=0]{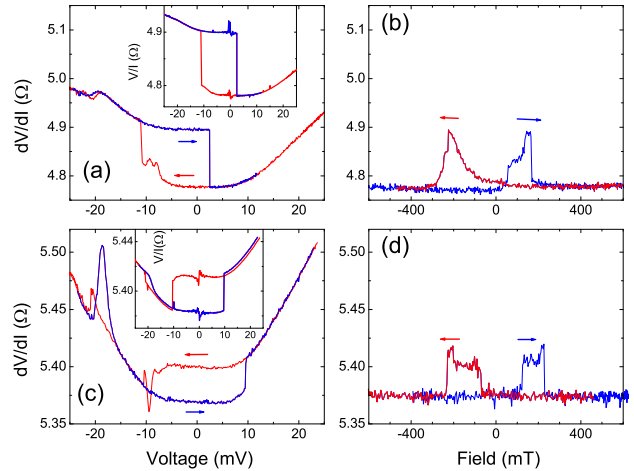}
\vspace{-1cm}
\end{center}
\caption{ Current (a) and field (b) driven \emph{normal} hysteresis
for a Cu30/Co5-Ag contact; (c) and (d) - same for a similar Cu30/Co5-Ag
contact exhibiting \emph{anomalous} hysteresis.}
\label{f3}
\end{figure}

Fig.\,3 illustrates the normal (a) and inverse (c) hysteresis in $R(V)$
accompanied by a spin-valve like hysteresis in $R(H)$ (b,d) for Cu30/Co5-Ag
contacts. Here the switching AP-to-P and P-to-AP occur as sharp steps
in resistance, occurring under the influence of a nominally unpolarized
current (insets a and c) or an externally applied magnetic field.
Importantly, the resistance difference between the stable at zero
bias AP and P states is the same for the current and field induced
transitions, even though the mechanisms leading to hysteresis are
different in the two cases. For $R(V)$ the bi-stability is caused
by spin transfer torques while for $R(H)$ with the external magnetic
field parallel to the film plane the behavior is identical to the
nano-pillar spin-valve magnetoresistance. This observation is true
even for $R(V)$ curves with smooth transitions such as Fig.\,2a,c
- the difference in resistance versus current and that versus field
are the same. Notice also the maxima in $dV/dI$ at approximately
$-20$ mV, which are due to non-hysteretic magnetization excitations
\cite{Ji,Yanson,Polianski}.

Another important observation, based on our measurements of about
one hundred contacts with hysteresis, is that statistically $\Delta R/R$
shows no trend as a function of the thickness of the Co layer in the
range $t=$2-100 nm, as illustrated in Fig.\,4a, strongly suggesting
that the observed hysteresis is a surface effect. As shown in Figs.
2,3,4b this surface effect can occur at either of the ferromagnetic
interface and is indistinguishable in its current and field driven
behavior from the standard three-layer spin-valve effect.

The above clear correlation between the dominant phonon maxima
and the sign of hysteresis in the conductance of the hetero-contacts
is straightforward conceptually, and reflects the limiting cases of
the ferromagnetic layer being at the bottom or top of the contact
core. Naturally, a richer behavior should be expected for very thin
magnetic layers located symmetrically in the PC core. Namely, the
spin torque effects from the two interfaces should superpose. An example
of such behavior is shown in Fig.\,4b. Here the polarity of the hysteresis
near zero bias is \emph{normal}, correlating well with the pronounced
Au phonon maximum at $\approx$10 mV. At positive bias, expected to
drive the contact into a low resistance state if one considers the
top interface only, such a \emph{normal} hysteresis transition indeed
occurs at $\approx$15 mV. However, the presence of the bottom interface
within the contact core, as evidenced by the Cu-phonon maximum in
the PC spectrum, results in a superposed \emph{anomalous} hysteresis
at large positive bias. The resulting structure of the ferromagnetic
layer of only 2 nm in thickness is a rather surprising double spin-valve
of type $\downarrow\uparrow\uparrow$ or $\uparrow\uparrow\downarrow$.
The data in Fig.\,2,3 with hysteresis of one type only represent therefore
the limiting cases where the PC conductance is dominated by either
one of the two N/F interfaces.

\begin{figure}[b]
\vspace{2cm}
\begin{center}
\includegraphics[width=9.5cm,angle=0]{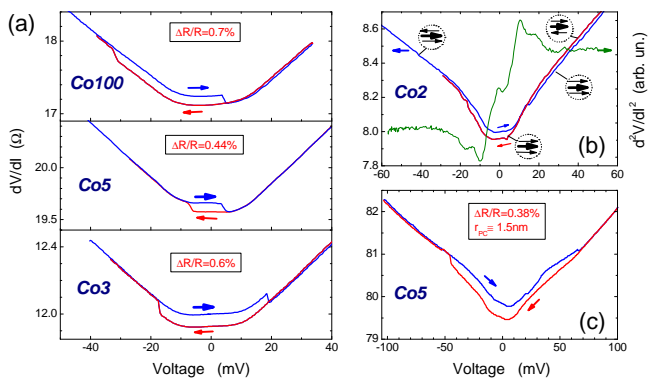}
\vspace{-4cm}
\end{center}
\caption{ (a) Hysteresis in $dV/dI(V)$ for contacts to Co films
of varying thickness; (b) $dV/dI(V)$ for a Cu100/Co2/Au3-Au contact
showing a superposed hysteresis from the Au and Cu interfaces and
the PC spectrum (green) exhibiting well defined phonon maxima for
both Au and Cu at $\approx10$ and $\approx17$ mV, respectively.
The arrows in dashed circles schematically indicate the orientation
of the surface and interior spins; (c) $dV/dI(V)$ for a high $R$
contact to a 5 nm Co film, $r_{\textrm{PC}}\approx1.5$ nm.}
\label{f3}
\end{figure}

The contact core, dominating the charge and spin current in PC's,
approaches 1 nm scale for highly resistive contacts with $R\sim100\ \Omega$.
$R(V)$ for such a contact, exhibiting a pronounced current driven
hysteresis, is shown in Fig.\,4c for a 5 nm thick Co layer. We estimate
the radius of this contact core to be $\approx1.5$ nm. Therefore
the whole F1-DW-F2 magnetic sub-layer structure producing the hysteretic
switching must be limited in thickness to some 3-4 atomic unit cells
at the interface, which provides a decisive evidence that the surface
spin layer acting as the {}``free'' layer is atomically thin. This
consideration and the lateral extent of the contact of only $\approx$3
nm (to an exchange-length thin ferromagnetic film) disqualifies any
interpretation of the phenomenon based on volume like domains or vortex
states. In particular, vortex states that are possible to produce
in $\sim100$ nm ferromagnetic particles in nanopillars should be
unstable for contacts to continuous films studied here, especially
at zero bias where circular Oersted fields are absent. We therefore
rule out the vortex interpretation for our contacts, which are one
to two orders of magnitude smaller than typical magnetic nanopillars.

The previously proposed interpretation \cite{Chen} of the \emph{normal}
hysteresis as due to a surface exchange bias in naturally oxidized
Co films is faced with difficulties as well. First of all, making
a true metallic contact removes any oxide from the contact region.
Furthermore, in contrast to \cite{Chen} all our measurements are
done on samples cooled in zero field, which we find does not diminish
the hysteresis effects. As shown in Fig.\,4b hysteresis is also observed
in contacts to ferromagnetic films capped with protective Au anti-oxidation
layers (we have measured a number of such contacts). As shown in 
Fig.\,2,3,4b we also detect hysteresis due to the bottom ferromagnetic interface,
which is produced in high vacuum and is certainly free from any oxide.
Finally, surface exchange anisotropy anyway necessitates postulating
a volume like domain \cite{Chen}, which can be excluded for our nano-contacts
to nanometer thin films. We therefore can rule out this mechanism
for the effects we observe. It is important to mention that we observe
pronounced magnetic hysteresis effects on Permalloy (Ni$_{80}$Fe$_{20}$)
films, which is an additional evidence against interpretations based
on surface exchange or stress induced anisotropy.

Spins at ferromagnetic interfaces can have substantially
different magnetic character and be weakly coupled to the interior
spins, which finds support in the recent studies of surface and interface
magnetism in Co and Fe \cite{Gruyters,Jonker}. The fundamental physics
involved is that the interface spins have a lower coordination number
and therefore fewer exchange bonds compared to the bulk spins. This
can lead to a reconstruction of the interface spin order, resulting
in quite different magnetic moment and anisotropy. The spin transport
effect we observe is of quite general nature and should also be present
in magnetic structures on a larger scale, provided that surface imperfections
such as roughness do not lead to averaging out its contribution. With
a proper control of the magnetic interface in nanodevices this effect
can offer a new way of manipulating the electron spin.

Acknowledgment. Financial support from the Swedish
Foundation for Strategic Research (SSF), the Royal Swedish
Academy of Sciences (KVA), and the National Academy
of Sciences of Ukraine (NASU) under project NANO is
gratefully acknowledged.

\end{document}